\documentstyle[preprint,aps,epsfig,floats]{revtex}
\tightenlines

\begin{document}

\title{Finite-size scaling in the steady state of the fully asymmetric
exclusion process \\}

\author{Jordan Brankov}
\address{Institute of Mechanics, Bulgarian Academy of Sciences, \\
Acad. G. Bonchev St. 4, 1113 Sofia, Bulgaria \\
e-mail: brankov@bas.bg}

\date{submitted to the Phys. Rev. E}
\date{\today}
\maketitle

\begin{abstract}
Finite-size scaling expressions for the current near the
continuous phase transition, and for the local density near the first-order
transition, are found in the steady state of the one-dimensional fully
asymmetric simple-exclusion process (FASEP) with open boundaries and
discrete-time dynamics. The corresponding finite-size scaling
variables are identified as the ratio of the chain length to the localization
length of the relevant domain wall.
\end{abstract}

\pacs{PACS numbers: 05.60.-k, 02.50.Ey, 05.70.Ln, 64.60.Ht}

\newpage

\section{Introduction}

We consider the fully asymmetric simple-exclusion process (FASEP) on a
finite chain of $L$ sites with open boundaries. For mathematical definition of
the exclusion processes we refer the reader to the book \cite{L99}, and for a
recent review on the relevant class of exactly solvable models for many-body
systems far from equilibrium to \cite{S01}, see also \cite{H00}. We recall
that each site $i \in \{1,2, \dots ,L\}$ of the chain is either empty or
occupied by exactly one particle.
The particles obey a discrete-time stochastic dynamics according to which
they hop with probability $p$ only to empty nearest-neighbor sites to the
right. The open boundary conditions imply that at each time step (update of
the whole chain) a particle is injected with probability $\alpha$ at the left
end of the chain ($i=1$), and removed with probability $\beta$ at the right
end ($i=L$). The order in which the local hopping, injection and particle
removal takes place is specified by one of the basic discrete-time updates,
see \cite{RSSS}. Here we explicitly consider the case of forward-ordered 
sequential update.	    

We mention that the case of random-sequential update was solved first by using
the recursion relations method \cite{DDM}, \cite{SD}, and then by means of the
matrix-product Ansatz (MPA) \cite{DEHP}. Next, the method of MPA was 
successfully applied for obtaining the steady-state properties in all the
basic cases of true discrete-time dynamics: forward- ($\rightarrow$) and
backward-ordered ($\leftarrow$) sequential \cite{RSS}, \cite{HP}, \cite{BPV},
sublattice-parallel ($\mbox{s-}\parallel$) \cite{H}, \cite{RS},
and, finally, fully parallel dynamics \cite{ERS}, \cite{GN}.

The phase diagram for all the discrete-time updates has the same structure, as
shown in Fig. 1: it
contains a maximum-current (m.c.), low-density (l.d.), and high-density (h.d.)
phases. The maximum-current phase is separated by lines of continuous
phase transitions, $(\alpha =\alpha_c,\; \beta_c \leq \beta \leq 1)$ and
$(\beta = \beta_c, \; \alpha_c \leq \alpha \leq 1)$, from the low-density and
high-density phases, respectively. Here $\alpha_c$ and $\beta_c$ are the
critical values of the injection and removal probabilities:
\begin{equation}
\alpha_c = \beta_c = 1-\sqrt{1-p}.
\label{acba}
\end{equation}          
The above phases were identified with respect to the analytic form of the bulk
current: for fixed $p$, the current in the low- (high-) density phase depends
only on $\alpha$ ($\beta$), and in the maximum-current phase it is independent
of both $\alpha$ and $\beta$. On crossing the borderline between the
maximum-current and the low- (high-) density phase, the current itself and its
first derivative with respect to $\alpha$ ($\beta$) change continously, and
the second derivative with respect to $\alpha$ ($\beta$) udergoes a finite
jump. The coexistence line between the low- and high-density phases is given
by $(\alpha =\beta , \; 0\leq \beta \leq \beta_c)$; on crossing it the bulk
density undergoes a finite jump.

Here, the exact finite-size expressions for the current and local density in
the steady state of FASEP with open boundaries and forward-ordered sequential
update, derived in \cite{BPV}, are analysed within the framework of
finite-size scaling (FSS) at continuous (for the current) and first-order (for
the density) phase transitions. The appropriate scaling variables are
identified and the corresponding scaling functions for the current and local
desity are explicitly obtained. The notions of
the Privman-Fisher anisotropic FSS have been recently extended to
non-equilibrium systems belonging to the directed percolation
and diffusion-annihilation universality classes, \cite{H00}, \cite{HS},
\cite{HH}. To the best of our knowledge, the present study
is the first step in the analytic confirmation of FSS for an exactly solved
model of a driven lattice gas with open boundaries. Since we are dealing
with phase transitions in the steady-state of the FASEP with discrete-time
updates, the equvalent two-dimensional lattice model is infinite in the
temporal direction but finite in the spatial one.

Note that the current has the same value in the cases of forward-ordered
($\rightarrow$), backward-ordered ($\leftarrow$) sequential, and
sublattice-parallel ($\mbox{s-}\parallel$) dynamics, i.e.,
$J_L^{\rightarrow}=J_L^{\leftarrow}= J_L^{\mbox{s-}\parallel}$. The
corresponding local densities at site $i \in \{1,\dots ,L\}$ are 
related to each other \cite{RS}:
\begin{equation}
\rho_L^{\rightarrow}(i)=\rho_L^{\leftarrow}(i)-J_L^{\rightarrow},\qquad
\rho_L^{\mbox{s-}\parallel}(i)= \left\{ \begin{array}{ll}
\rho_L^{\rightarrow}(i),&\quad \mbox{$i$ odd} \\
\rho_L^{\leftarrow}(i),&\quad \mbox{$i$ even} .\end{array} \right.
\label{slor}
\end{equation}
As shown in \cite{ERS}, the current $J_L^{\parallel}$ and local
density $\rho_L^{\parallel}(i)$ for the FASEP with fully parallel update can
be simply expressed in terms of those for the forward-ordered sequential
update:
\begin{equation}
J_L^{\parallel}={J_L^{\rightarrow}\over 1+J_L^{\rightarrow}} , \qquad
\rho_L^{\parallel}(i)={\rho_L^{\rightarrow}(i) +J_L^{\rightarrow}\over
1+J_L^{\rightarrow}}.
\label{m1}
\end{equation}
Due to these relations, the results derived here suffice to explicitly obtain
the current and density FSS functions for each of the basic discrete-time
updates.
 
Concerning the notation, we note that the exact finite-chain results obtained
in \cite{BPV} are conveniently expressed in terms of the parametrs 
\begin{equation}
d = \sqrt{1-p},\quad a=d+d^{-1}, \quad \xi ={p-\alpha \over \alpha d},
\quad \eta = {p-\beta \over \beta d},
\label{dke}
\end{equation}
which will be used here too.

\section{Finite-size scaling at the continuous phase transition}

Let us consider first the continuous phase transition across the boundary
$\alpha =\alpha_c, \; \beta_c \leq \beta \leq 1$ between the low-density
phase and the maximum-current phase. In terms of the variables (\ref{dke}) the 
equation of this boundary reads $\xi=1,\; -d \leq \eta \leq 1$; the m.c.
phase occupies the region $-d \leq \xi \leq 1,\; -d \leq \eta \leq 1$, and the
region $\xi >1 \geq \eta \geq -d$, called region AII, lies in
the l.d. phase, see Fig. 1.

According to the basic FSS hypotheses, the FSS variable in the case of a
continuous transition, characterized by diverging bulk correlation length
$\lambda$, should be given by the ratio $L /\lambda$, where $L$ is the
finite-size of the system. As it is well known, in the case at hand
the inverse correlation length $\lambda_{\xi}^{-1}$ in the l.d. phase is,
see, e.g., \cite{GN}, \cite{BP},
\begin{equation}
\lambda_{\xi}^{-1} = \ln \left[1 +{(\xi -1)^2\over \xi (a+2)}\right],
\quad (\xi \geq 1),
\label{cordt}
\end{equation}
and $\lambda_{\xi}^{-1} \equiv 0$ in the m.c. phase.
Hence, the FSS variable as $L\rightarrow \infty$ and $\xi \rightarrow 1^+$
is expected to be given by
\begin{equation}
L/\lambda_{\xi} \simeq {L(\xi -1)^2 \over \xi (a+2)} := x_1^2(L,t),
\label{fssvar1}
\end{equation} 
where
\begin{equation}
x_1(L,t) \simeq C_1 t L^{1/2},\quad C_1 =(a+2)^{-1/2}, \quad t= \xi -1.
\label{var1}
\end{equation} 
We emphasize that here we study a boundary induced non-equilibrium phase
transition, and the physical quantity which measures the distance from the
steady-state critical point is related to the injection probability:
$t= \xi -1 \sim 1-\alpha /\alpha_c$.

Consider now the finite-size current $J_L(\xi,\eta)=Z_{L-1}(\xi,\eta)/
Z_L(\xi,\eta)$, where $Z_L(\xi,\eta)$ is the normalization constant of the
steady-state probability for a lattice of $L$ sites. The following exact
representation of $Z_L(\xi,\eta)$ in the subregion AII of the low-density
phase ($\xi >1 \geq \eta$) has been found in \cite{BPV}:
\begin{equation}
\label{51BC}
Z_L^{\rm AII}(\xi,\eta)=\left({d\over p}\right)^L{\xi -\xi^{-1}\over \xi-\eta}
(a +\xi +\xi^{-1})^L + Z_L^{\rm m.c.}(\xi, \eta).
\end{equation}
Here the expression for the normalization constant in the maximum-current
phase ($\xi \not= \eta$),
\begin{equation}
\label{51}
Z_L^{\rm m.c.}(\xi,\eta)= \left({d\over p}\right)^L \left[{\xi\over \xi-\eta}
I_L(\xi) + {\eta \over \eta -\xi}I_L(\eta)\right],
\end{equation}
involves the integral
\begin{equation}
\label{51n}
I_L(\xi)= \frac{2}{\pi} \int_0^{\pi} {\rm d} \phi {(a+2
\cos\phi)^L \sin^2 \phi \over 1-2\xi\cos \phi +\xi^2} ,
\end{equation}
which is a non-analytic function of $\xi$ at $\xi =1$. For all finite $L$ the
normalization constant $Z_L^{\rm AII}(\xi,\eta)$ in region AII 
represents an analytic continuation of $Z_L^{\rm m.c.}(\xi,\eta)$ from the
domain $|\xi|<1$ to the domain $|\xi| >1$, see \cite{BPV}.

A direct application of Laplace's method for evaluation of the integral
(\ref{51n}) as $L\rightarrow \infty$ shows that it changes its
leading-order asymptotic
behavior from $O(L^{-3/2})$ for $\xi \not= 1$ to
$O(L^{-1/2})$ for $\xi = 1$, see Eq. (14) in \cite{BP}. The finite-size
expression that interpolates between these limiting asymptotic forms can be
readily obtained by using small-argument expansion of the trigonometric
functions in the integrand. The result for $x_1(L,t)=O(1)$ is
\begin{equation}
\label{51ss}
I_L(\xi)\simeq {(a+2)^{L+1/2}\over \xi \sqrt{\pi L}} X(|x_1|),
\end{equation}
where the FSS function $X(\cdot)$ is given by
\begin{equation}
\label{X}
X(x) = 1 -\sqrt{\pi}x{\rm e}^{x^2}[1-\Phi (x)],
\quad \Phi (x)={2\over \sqrt{\pi}}\int_0^x {\rm e}^{-t^2}{\rm d}t.
\end{equation}
Thus, keeping the $1/L$ corrections to the finite-size scaling form, we
obtain
\begin{equation}
\label{Zfs}
Z_L^{\rm m.c.}(\xi,\eta)\simeq \left({d\over p}\right)^L {1\over \xi - \eta}
{(a+2)^{L+1/2}\over \sqrt{\pi L}}\left[X(|x_1|)- {\eta \over (1-\eta)^2}
{a+2 \over 2L}\right].
\end{equation}
The small- and large-argument asymptotic behavior of $X(\cdot)$ readily
follows from that of the Fresnel integral $\Phi (\cdot)$:
\begin{equation}
\label{Xas}
X(x) = \left\{\begin{array}{ll} 1 -\sqrt{\pi}x +O(x^2),\quad &\mbox{\rm as}
\quad x\rightarrow 0, \nonumber \\
{1\over 2}x^{-2} -{3\over 4}x^{-4} +O(x^{-6}),\quad &\mbox{\rm as}
\quad x\rightarrow \infty.\end{array} \right.
\end{equation}

Let us first evaluate the asymptotic behvior of the finite-size correction to
the bulk current in the maximum-current phase, 
\begin{equation}
\label{fsop1mc}
J_L^{\rm m.c.}(\xi,\eta)-J_{\infty}^{\rm m.c.}=
{Z_{L-1}^{\rm m.c.}(\xi,\eta)\over Z_L^{\rm m.c.}(\xi,\eta)}-
{p\over d(a+2)},
\end{equation}
as $L\rightarrow \infty$ and $\xi \rightarrow 1^-$ at fixed $\eta <1$.
We readily obtain in the leading order of magnitude,
\begin{eqnarray}
\label{id1mc}
\xi [(a+2)I_{L-1}(\xi)-I_L(\xi)]&\simeq &{|1-\xi|^{3/2}\over \xi^{3/2}}
{2\over \pi}\int_0^{\infty} {\rm d}x {{\rm e}^{-x_1^2 x^2}x^4 \over 1+x^2}
\nonumber \\ &=& {(a+2)^{3/2}\over 2\sqrt{\pi} L^{3/2}}[1-2x_1^2X(|x_1|)].
\end{eqnarray}
Hence we derive
\begin{equation}
\label{JLJbulk}
J_L^{\rm m.c.}(\xi,\eta)- J_{\infty}^{\rm m.c.}\simeq {1\over L}
{p\over 2d(a+2)}\left[{1- 2x_1^2 X(|x_1|)\over X(|x_1|)}\right].
\end{equation}
The asymptotic behavior of the above finite-size correction to the current as
$x_1\rightarrow 0^-$ (say, as $\xi\rightarrow 1^-$ at large fixed $L$) follows
from Eq. (\ref{Xas}):
\begin{equation}
\label{fsop1f1mc}
J_L^{\rm m.c.}(1,\eta)-J_{\infty}^{\rm m.c.}
\simeq {1\over L}{p\over 2d(a+2)}.
\end{equation}
In the limiting case $x_1\rightarrow -\infty$ (say, $L\rightarrow \infty$
at $\xi <1$ fixed close to unity), we obtain that the finite-size correction
is again of the order $O(L^{-1})$:
\begin{equation}
\label{fsop1f2}
J_L^{\rm m.c.}(\xi,\eta)-J_{\infty}^{\rm m.c.}\simeq {1\over L}{3p\over
2d(a+2)}.
\end{equation}

Consider next the asymptotic behvior of the finite-size correction to the
bulk current in the low-density phase, 
\begin{equation}
\label{fsop1}
J_L^{\rm AII}(\xi,\eta)-J_{\infty}^{\rm l.d.}(\xi)=
{Z_{L-1}^{\rm AII}(\xi,\eta)\over Z_L^{\rm AII}(\xi,\eta)}-
{p\over d(a+\xi +\xi^{-1})},
\end{equation}
as $L\rightarrow \infty$ and $\xi \rightarrow 1^+$ at fixed $\eta <1$.
With the aid of the identities:
\begin{equation}
\label{id1}
\xi [(a+\xi +\xi^{-1})I_{L-1}(\xi)-I_L(\xi)]=I_{L-1}(0),
\end{equation}
\begin{equation}
\label{id2}
\eta [(a+\xi +\xi^{-1})I_{L-1}(\eta)-I_L(\eta)]=I_{L-1}(0)+
(\xi -\eta)(\eta -\xi^{-1})I_{L-1}(\eta),
\end{equation}
we obtain in the leading order:
\begin{equation}
\label{fsop1f}
J_L^{\rm AII}(\xi,\eta)-J_{\infty}^{\rm l.d.}(\xi)\simeq {1\over L}
{p\over 2d(a+2)}\left[{1\over 2\sqrt{\pi}x_1{\rm e}^{x_1^2}+X(x_1)}\right] .
\end{equation}
Hence, by using Eq. (\ref{Xas}) we readily derive the asymptotic behavior of
the above finite-size correction as $x_1\rightarrow 0^+$ (say, as $\xi
\rightarrow 1^+$ at large fixed $L$): 
\begin{equation}
\label{fsop1f1}
J_L^{\rm AII}(1,\eta)-J_{\infty}^{\rm l.d.}(1)\simeq {1\over L}{p\over 2d(a+2)}.
\end{equation}
In the other limiting case $x_1\rightarrow \infty$ (say, $L\rightarrow \infty$
at $\xi >1$ fixed close to unity), we obtain that the finite-size corrections
to the current are exponentially small:
\begin{equation}
\label{fsop1f3}
J_L^{\rm AII}(\xi,\eta)-J_{\infty}^{\rm l.d.}(\xi) = O\left(x_1^{-1}
{\rm e}^{-x_1^2}\right).
\end{equation}

As a finite-size order parameter of the continuous phase transition
one could consider the difference in the finite-size currents:
\begin{equation}
\label{fsop2}
\Delta_L(\xi,\eta):= J_L^{\rm m.c.}(\xi,\eta)-J_L^{\rm AII}(\xi,\eta)
\quad (\xi >1, \eta <1).
\end{equation}
Here $J_L^{\rm m.c.}(\xi,\eta)$ represents the analytic expression for
the finite-size current in the maximum-current phase, i.e., $J_L^{\rm m.c.}=
Z_{L-1}^{\rm m.c.}/Z_L^{\rm m.c.}$ with $Z_L^{\rm m.c.}$ defined by
Eqs. (\ref{51}) and (\ref{51n}), evaluated in subregion AII of the low-density
phase (at $\xi >1, \eta <1$). For the corresponding bulk quantity we have
\begin{equation}
\label{bulkop}
\Delta_{\infty}(\xi) = J_{\infty}^{\rm m.c.}-J_{\infty}^{\rm l.d.}(\xi)=
{(\alpha -\alpha_c)^2\over p(1-\alpha)},
\end{equation}
which suggests the critical exponent for the order parameter $\beta =2$.
In the finite-size case, by taking into account that
\begin{equation}
\label{trivial}
{p\over d(a+2)}-{p\over d(a+\xi +\xi^{-1})}= {1\over L}
{p\over d(a+2)}{x_1^2 \over 1+x_1^2 /L},
\end{equation}
and combining the above results, we obtain in the leading order
\begin{equation}
\label{opf}
\Delta_L(\xi,\eta) \simeq {1\over L}{p\over 2d(a+2)}\left\{
{x_1 \over  X(x_1)\left[x_1+ {1\over 2\sqrt{\pi}}
{\rm e}^{-x_1^2}X(x_1)\right]}\right\}. 
\end{equation}

Since the high-density phase maps onto the
low-density phase under the exchange of arguments $\xi \leftrightarrow \eta$
(equivalently, $\alpha \leftrightarrow \beta$), the FSS properties of the
continuous transition across the boundary
$\beta =\beta_c, \; \alpha_c \leq \alpha \leq 1$ between the high-density
phase and the maximum-current phase follow trivially from the above results
and the particle-hole symmetry.

\section{Finite-size scaling at the first-order transition}

In the thermodynamic limit the first-order phase transition, which occurs
across the borderline $\beta =\alpha,\;0\leq \alpha \leq \alpha_c$
($\eta =\xi,\; \xi \geq 1$) between subregions AI and BI, manifests itself by
a finite jump in the bulk density:
\begin{equation}
\label{jump}
\left. \rho_{\rm bulk}^{\rm h.d.}(\eta) - \rho_{\rm bulk}^{\rm l.d.}
(\xi)\right|_{\xi =\eta}= {\eta -\eta^{-1}\over a+\eta +\eta^{-1}} >0.
\end{equation}
Quite peculiarly, this transition is characterised by another
{\em diverging} correlation length
\begin{equation}
\label{lambda}
\lambda^{-1} = |\lambda_{\xi}^{-1} - \lambda_{\eta}^{-1}| =
{1-\eta^{-2}\over a+\eta +\eta^{-1}}|\xi -\eta| +O(|\xi -\eta|^2),
\end{equation}
which suggests                                           the finite-size scaling variable
\begin{equation}
\label{fofssv}
\tilde{x_2}:= L/\lambda \simeq C_2|h|L, \quad C_2 =
{1-\eta^{-2}\over a+\eta +\eta^{-1}},\quad h= \xi -\eta.
\end{equation}

Our analysis starts with the exact result found for the finite-chain local density
$\rho_L^{\rm D}(i;\xi,\eta)$ in region D$=$AI$\cup$BI of the phase diagram,
see Eqs. (4.21) and (A13) in \cite{BPV}. This result can be cast in the form
($\xi \not= \eta$):
\begin{equation}
\label{rhoD}
\rho_L^{\rm D}(i;\xi,\eta)= {1-p\over p}J_L^{\rm D}(\xi,\eta)
+\widetilde{\Omega}_L^{\rm D}(i;\xi,\eta),
\end{equation}
where,
\begin{eqnarray}
\widetilde{\Omega}_L^{\rm D}(i;\xi,\eta)= 
\left(\frac{d}{p}\right)^L{(a+2)^{L-1}\over (\xi-\eta)Z_L^{\rm D}}\left[
(1-\xi^{-2}){\rm e}^{(L-1)/\lambda_{\xi}}
\right. \nonumber \\ \left. + (\xi-\xi^{-1})(\eta -\eta^{-1})
{\rm e}^{(i-1)/\lambda_{\xi}}{\rm e}^{(L-i)/\lambda_{\eta}}-
(\eta^2 -1){\rm e}^{(L-1)/\lambda_{\eta}}\right] \nonumber \\ 
-\left(\frac{d}{p}\right)^L{(a+2)^{L-1}\over Z_L^{\rm D}}\left[
(\xi-\xi^{-1}){\rm e}^{(i-1)/\lambda_{\xi}}\left({p\over d}\right)^{L-i}
{Z_{L-i}^{\rm m.c.}\over (a+2)^{L-i}}
\right. \nonumber \\ \left. -(\eta-\eta^{-1})
{\rm e}^{(L-i)/\lambda_{\eta}}\left({p\over d}\right)^{i-1}
{Z_{i-1}^{\rm m.c.}\over (a+2)^{i-1}}\right] \nonumber \\
+\frac{d}{2p}{1\over Z_L^{\rm D}}\left[F_L(i;\xi,\eta)-
(\xi -\eta)Z_{i-1}^{\rm m.c.} Z_{L-i}^{\rm m.c.} +
\frac{p}{d}Z_L^{\rm m.c.}-aZ_{L-1}^{\rm m.c.}\right] .
\label{y1}
\end{eqnarray}
For brevity of notation we have omitted the arguments $\xi$ and $\eta$ of the function
$Z_L(\xi,\eta)$. 
Here the term $F_L(i)=F_L(i;\xi,\eta)$ is the antisymmetric (with
respect to the center of the chain) function of the integer coordinate $i$,
$F_L(i;\xi,\eta)=-F_L(L-i+1;\xi, \eta)$, defined for $1\leq i \leq [L/2]$
(where $[x]$ denotes the integer part of $x$) by the equation
\begin{equation}
\label{72}
F_L(i;\xi,\eta)= \left({d\over p}\right)^{L-1}(1-\xi \eta)
\sum_{n=0}^{L-2i}I_{L-i-n-1}(\xi)I_{i+n-1}(\eta).
\end{equation}
The normalization constant in the region D$=$AI$\cup$BI ($\xi >1,\eta >1$)
for $\xi \not= \eta$ is given by
\begin{equation}
\label{51D}
Z_L^{\rm D}(\xi,\eta)=\left({d\over p}\right)^L{(a+2)^L\over \xi-\eta}
\left[(\xi-\xi^{-1}){\rm e}^{L/\lambda_{\xi}}-(\eta-\eta^{-1})
{\rm e}^{L/\lambda_{\eta}}\right] + Z_L^{\rm m.c.}(\xi,\eta).
\end{equation}
Let us analyse this expression when
$\xi = \eta+h$, $h\rightarrow 0$. For $Z_L^{\rm m.c.}(\eta +h,\eta)$ we have
\begin{eqnarray}
\label{51Dh}
Z_L^{\rm m.c.}(\eta +h,\eta)&=&\left({d\over p}\right)^L{1\over h}[(\eta +h)
I_L(\eta +h)-\eta I_L(\eta)]\nonumber \\
&=&-\left({d\over p}\right)^L(\eta^2 -1)K_L(\eta)+ O(h),
\end{eqnarray}
where
\begin{equation}
\label{51nD}
K_L(\eta)= \frac{2}{\pi} \int\limits_0^{\pi} {\rm d} \phi {(a+2
\cos\phi )^L \sin^2 \phi \over (1-2\eta\cos \phi +\eta^2)^2}.
\end{equation}
For $\xi = \eta +h$, as $L\rightarrow \infty$ and $h\rightarrow 0$ we have
\begin{equation}
\label{lambdaxi}
L/\lambda_{\eta +h}= L/\lambda_{\eta} + C_2 hL +O(Lh^2),
\end{equation}
hence
\begin{equation}
\label{ZDas}
Z_L^{\rm D}(\eta + h,\eta)=\left({d\over p}\right)^L(a+2)^L\left[
(\eta-\eta^{-1}){\rm e}^{L/\lambda_{\eta}}{{\rm e}^{C_2hL}-1\over h}
+O(1)\right].
\end{equation}
Next, by using the upper bound $I_k(\xi)\leq (a+2)^kI_0(\xi)$, where
$I_0(\xi)=1$ for $|\xi|\leq 1$ and $I_0(\xi)=\xi^{-2}$ for $|\xi|\geq 1$,
one easily obtains that in region D
\begin{equation}
\label{Fas}
\left|F_L(i;\xi,\eta)\right| \leq = \left({d\over p}\right)^{L-1}
{\xi \eta -1\over \xi^2 \eta^2}(L-1)(a+2)^{L-2}.
\end{equation}
Therefore, in view of Eq. (\ref{ZDas}), the contribution of the term
proportional to $F_L(i;\xi,\eta)$ into the local density is exponentially
small. Thus, only the first three terms in the right-hand side of
Eq. (\ref{y1}) contribute into the leading-order expression for
$\widetilde{\Omega}_L^{\rm D}(\eta + h,\eta)$:
\begin{equation}
\label{OMas}
\widetilde{\Omega}_L^{\rm D}(i;\eta +h,\eta)\simeq {\eta^{-1}\over a+\eta +
\eta^{-1}} + (\eta-\eta^{-1}){{\rm e}^{C_2 hi} -1\over {\rm e}^{C_2hL}-1}.
\end{equation}
In deriving the above expression we have assumed that both $i\gg 1$ and
$L-i \gg 1$. Finally, by taking into account that 
\begin{equation}
\label{Jas}
J_L^{\rm D}(\eta +h,\eta) = {p\over d(a+ \eta + \eta^{-1})} +O(h),
\end{equation}
we obtain  the leading order expression for the local density on the
macroscopic scale $i/L=r$, $0<r<1$:
\begin{equation}
\label{rhoDas}
\rho_L^{\rm D}(rL;\xi,\eta)= {1\over a+ \eta + \eta^{-1}}
\left[ d + \eta^{-1}+(\eta-\eta^{-1}){{\rm e}^{x_2 r} -1\over
{\rm e}^{x_2}-1}\right].
\end{equation}
Here we have introduced the FSS variable $x_2 = C_2 hL$, compare with
Eq. (\ref{fofssv}).
In the limit $x_2 \rightarrow 0$ the above expression reduces to the
well-known linear density profile on the coexistence line $\xi =\eta$:
\begin{equation}
\label{rholin}
\rho_L^{\rm D}(rL;\eta,\eta)= {d + \eta^{-1}+(\eta-\eta^{-1}) r
\over a+ \eta + \eta^{-1}}.
\end{equation}
In the limit $x_2 \rightarrow + \infty$ ($x_2 \rightarrow - \infty$) one
recovers the bulk density in the low-density (high-density) phase.

\newpage

\section{Discussion}

The mathematical mechanism of the phase transitions in the FASEP with open
boundaries has been revealed in \cite{BPV} as qualitative changes in the
spectrum of the lattice translation operator $C$ \cite{SS}. In the region
$\xi \leq 1$, $\eta \leq 1$, occupied by the maximum-current phase, the
spectrum is continuous and fills
with uniform density the interval from $(d/p)(a-2)$ to $(d/p)(a+2)$. When
$\eta \leq 1$ but $\xi$ becomes larger than unity, an eigenvalue
\begin{equation}
\label{eigen}
\nu (\xi) = (d/p)(a + \xi +\xi^{-1}) 
\end{equation}
splits up from the continuous spectrum and dominates the properties of the
low-density phase in region AII. Due to particle-hole symmetry, when
$\xi \leq 1$ but $\eta$ exceeds unity, the eigenvalue that splits up from
the continuous spectrum and dominates the properties of the high-density phase
in region BII is $\nu (\eta)$. Thus, the logarithm of the ratio of the
corresponding eigenvalue to the ceiling of the continuous spectrum defines the
relevant inverse correlation length $\lambda_{\xi}^{-1}$ or
$\lambda_{\eta}^{-1}$, see Eq. (\ref{cordt}). When both $\xi > 1$ and
$\eta > 1$, which is the case in region D$=$AI$\cup$BI, there are two
eigenvalues, $\nu (\xi)$ and $\nu (\eta)$, above the continuous spectrum.
Obviously, these eigenvalues become degenerate on the coexistence line
$\xi = \eta >1$, which explains the appearance of the diverging correlation
length (\ref{lambda}). Note that the explicit expressions for the correlation
lengths depends on the type of update only: they are the same for all the true
discrete-time updates, and for the random sequential update, see Eq. (78) in
\cite{DEHP},
\begin{equation}
\lambda_{\alpha}^{-1} = \ln \left[1+{(1/2-\alpha)^2\over \alpha (1-\alpha)}
\right].
\label{corct}
\end{equation} 

The physical meaning of the above correlation lengths emerges in the domain
wall picture developed in \cite{KSKS}: the lengths $\lambda_{\xi}$,
$\lambda_{\eta} $, and $\lambda$ are interpreted as localization lengths of
the domain walls between the low-density/maximum-current,
high-density/maximum-current, and low-density/high-density phases,
respectively. The complete delocalization of the low-density/high-density
domain wall on the coexistence line explains the linear density profile
(\ref{rholin}): it is the result of ensemble averaging over configurations
with uniform probability distribution of the domain wall position \cite{SD}.

Thus, the FSS variable for any of the phase transitions in the
FASEP with open boundaries has the physical meaning of a ratio of the chain
length to the localization length of the relevant domain wall. 
The explicit expressions for the FSS functions have been derived 
here for the model with forward-ordered sequential update. The 
corresponding expressions for the other basic discrete-time updates 
follow under the mappings mentioned in the Introduction.

\section*{Acknowledgments}

The author thanks Dr. D. M. Danchev for stimulating discussions.
He is grateful to the International Atomic Energy Agency and UNESCO for
hospitality at the Abdus Salam International Centre for Theoretical Physics,
Trieste, where this work was started.

\newpage

\begin{figure}
\epsfxsize=15cm
\epsfysize=15cm
\epsfig{file=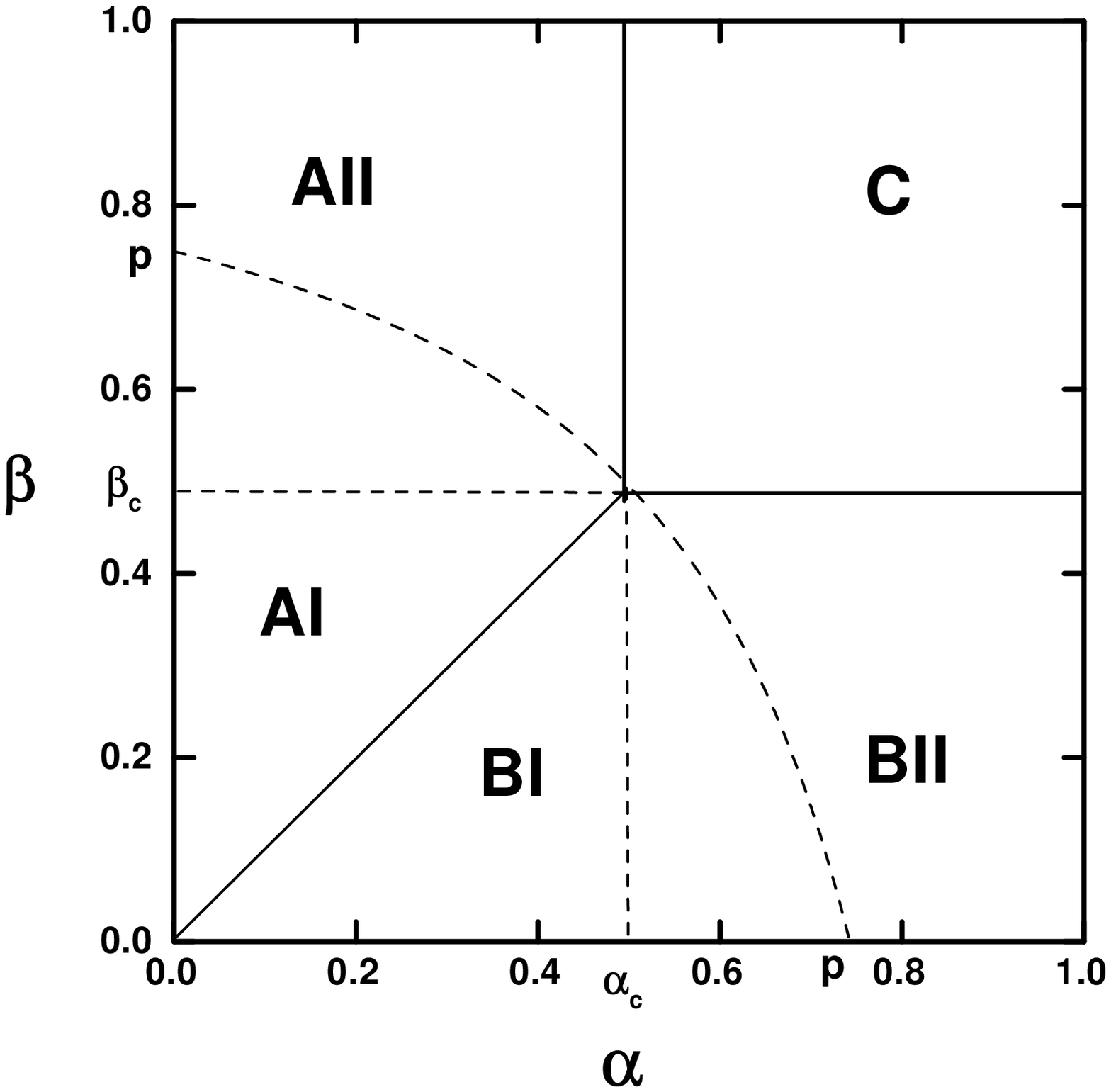}
\caption{The phase diagram in the plane of the injection and removal
probabilities $\alpha$ and $\beta$ (see the text) for hopping probability
$p$=0.75. The maximum-current phase occupies region C. Region
A$=$AI$\cup$AII corresponds to the low-density phase, 
and region B$=$BI$\cup$BII to the high-density phase.
Subregions AI (BI) and AII (BII) are distinguished 
by the different analytic form of the density profile. The boundary between
them, $\beta = \beta_c$, $0\leq \alpha \leq \alpha_c$
($\alpha = \alpha_c$, $0\leq \beta \leq \beta_c$), is shown by dashed
segment of a straight line. The solid line $\alpha = \beta$ between 
subregions AI and BI is the coexistence line of the low- and high-density
phases. The curved dashed line is the mean-field line $(1-\alpha)(1-\beta) =
1-p$.}
\label{fig:1}
\end{figure}

\end{document}